%% file: Interference_Main.tex
\documentclass[10pt,conference]{IEEEtran}

\usepackage{balance}
\usepackage{comment}
\usepackage{amsmath,amssymb,amsthm,amsfonts}
\usepackage{mathtools}
\usepackage{bigints}
\usepackage[printonlyused]{acronym}
\usepackage{nameref}
\usepackage{prettyref}
\usepackage{nicefrac}
\usepackage{afterpage}
\usepackage{algorithm}
\usepackage{algpseudocode}
\usepackage[pdftex]{graphicx}
\graphicspath{{./figures/}}
\usepackage{eqparbox}
\usepackage[usenames,dvipsnames]{xcolor}
\usepackage{cite}
\usepackage[normalem]{ulem}
\usepackage{soul}
\usepackage[footnote,draft,silent,nomargin]{fixme}
\usepackage{array}

\ifCLASSOPTIONcompsoc
\usepackage[caption=false,font=normalsize,labelfont=sf,textfont=sf]{subfig}
\else
\usepackage[caption=false,font=footnotesize]{subfig}
\fi

\newcommand{\M}{\mathcal{M}}

\newcommand{\Kd}{\mathcal{K}}

\def\conv{\otimes}

\def\eps{\varepsilon}
\def\minplus{{$(\min, +)\,$}}
\def\S{{\cal S}}
\def\A{{\cal A}}
\def\D{{\cal D}}

\def\M{{\mathcal M}}

\def\mx{{$(\min, \times)$}}

\newtheorem{theorem}{Theorem}
\newtheorem{lemma}{Lemma}
\newtheorem{corollary}{Corollary}

\newcounter{tempEquationCounter}
\newcounter{thisEquationNumber}

\allowdisplaybreaks

\IEEEoverridecommandlockouts 

\begin{document}

\title{On the Delay Performance of Interference Channels}

\author{\IEEEauthorblockN{\quad}
\IEEEauthorblockA{\quad\\ \quad\\ \quad}}

\author{
   \IEEEauthorblockN{Sebastian Schiessl\IEEEauthorrefmark{4}, Farshad Naghibi\IEEEauthorrefmark{4}, Hussein Al-Zubaidy\IEEEauthorrefmark{4}, Markus Fidler\IEEEauthorrefmark{3}\thanks{This work was supported in part by the European Research Council (ERC) under Starting Grant UnIQue (StG 306644).}, James Gross\IEEEauthorrefmark{4}}
   \IEEEauthorblockA{\IEEEauthorrefmark{4}School of Electrical Engineering, KTH Royal Institute of Technology, Stockholm, Sweden}
   \IEEEauthorblockA{\IEEEauthorrefmark{3}Institute of Communications Technology, Leibniz Universit\"at Hannover, Germany}
   Emails:$\{$schiessl,farshadn,hzubaidy,jamesgr$\}$@kth.se, and fidler@ieee.org
}
\maketitle

\input{Abstract}

\input{Introduction}

\input{Preliminaries}

\input{Derivations}

\input{Evaluations}

\input{Conclusions}

\balance
\bibliographystyle{IEEEtran}
\bibliography{Interference_Main}

\end{document}

%% file: Abstract.tex
\begin{abstract}
A deep understanding of the queuing performance of wireless networks is essential for the advancement of future wireless communications. The stochastic nature of wireless channels in general gives rise to a time varying transmission rate. In such an environment, interference is increasingly becoming a key constraint. Obtaining an expressive model for offered service of such channels has major implications in the design and optimization of future networks. However, interference channels are not well-understood with respect to their higher layer performance. The particular difficulty for handling interference channels arises from the superposition of random fading processes for the signals of the transmitters involved (i.e., for the signal of interest and for the signals of the interferers). Starting from the distribution of the signal-to-interference-plus-noise ratio (SINR), we derive a statistical characterization of the underlying service process in terms of its Mellin transform. Then, we adapt a recent stochastic network calculus approach for fading channels to derive measures of the queuing performance of single- and multi-hop wireless interference networks. Special cases of our solution include noise-limited and interference-limited systems. A key finding of our analysis is that for a given average signal and average sum interference power, the performance of interfered systems not only depends on the relative strength of the sum interference with respect to the signal-of-interest power, but also on the interference structure (i.e., the number of interferers) as well as the absolute levels. 
\end{abstract}

%% file: Introduction.tex
\section{Introduction}
\label{sec:introduction}
Over the last decade interference has become the key bottleneck for the further evolution of wireless networks.
With the advent and proliferation of broadband wireless communication services, this interference limitation is apparent in multiple ways.
For unlicensed bands, the interference limitation is due to the constantly increasing number of deployed wireless systems, running heterogeneous technologies and not undergoing a deployment planning.
This has led to quite crowded frequency bands which are facing a significant coexistence problem~\cite{Mass_Consult_09}.
On the other hand, the need for higher communication rates has forced cellular network providers to operate advanced packet-switched networks with a frequency reuse of one, i.e., potentially introducing a significant interference coupling between neighboring cells of the same system.
Consequently, (mitigating) the impact of interference in wireless communication networks has become an intense area of research recently~\cite{Cadambe_2008,Gesbert_2010}.
Despite the large research interest with respect to interference channels on the physical layer, little is known with respect to the impact of interference regarding the higher layers.
In particular, from a fundamental point of view only few attempts have been made to characterize the interference channel from a queuing-theoretic perspective.
Hence, models to study the performance of wireless (interfered) store-and-forward networks are lacking.

For instance, a higher-layer queuing analysis of wireless networks under the impact of interference is presented in~\cite{Zhang_2013} for sensor networks.
However, the considered interference model relates to the so-called protocol model, where nodes avoid interference for example based on CSMA/CA coordination.
Furthermore, the authors studied only average performance metrics by mapping the transmission behavior to standard G/G/1 queuing models.
A related analysis in the context of multi-hop networks is presented in~\cite{Bisnik2009}.
Here, the authors also study the protocol interference model and analyze multi-hop packet transmissions based on the G/G/1 open queuing network model, providing average performance metrics like the delay.
However, the more subtle effects of interference on the physical layer due to fading are not taken into account, while the analysis also falls short of providing a characterization of the end-to-end delay distribution.
The concept of effective capacity can address these issues, in particular it provides bounds on the tail of the delay of a single-hop communication systems.
However, typically interference is characterized in this context only as an additional constant contribution to the noise, ignoring the fading characteristic of interference~\cite{Musavian_2010,Elalem_2013}.
Finally,~\cite{Le_10} analyzes the interference channel with respect to scheduling stability in larger ad-hoc networks under the well-known Lyapunov stability framework.
While stability is an important aspect of queuing networks, further relevant metrics like the delay distribution cannot be addressed by this analysis.

In summary, most of the above works make a significant contribution towards understanding the average queuing performance of wireless networks under interference.
However, they cannot provide a more fine-grained analysis especially of the delay distributions.
Furthermore, a common assumption among these works is the transmission of a single constant rate traffic stream over a single interference channel.
The characterization of multi-hop performance of interference channels for variable rate traffic streams, in particular with respect to delay distributions, is an open challenge. This is especially true when it comes to precise models of the physical layer that also take into account the fading of the interference signals.
This clearly limits these approaches in terms of their expressibility with respect to fading profiles or transmit power settings.

In this work, we provide a network-layer performance analysis of interference channels in terms of their fading parameters. To our knowledge, this aspect has not been addressed before.
To enable such analysis, we utilize recent results from the literature, namely (i) the fading distribution of interference channels \cite{parruca_2015}, and (ii) the \mx~network calculus for wireless network analysis \cite{alzubaidy}.
A key step in such analysis is the characterization of the service element (in this case the interference channel) which then enables the determination of the desired performance bounds.
Our main contribution is to provide such mathematical characterization of the service offered by the interference channel, i.e., a fading channel in the presence of multiple (fading) interferers, in terms of the Mellin transform of the cumulative service process of the channel.
Computing this particular Mellin transform involves the solution of an integral transform of a \textit{ratio distribution}  which is known to be notoriously difficult to handle.
{We then use the resulting service process to obtain probabilistic bounds on the delay performance and provide
the corresponding delay performance for some special cases.}

This contribution has four main and novel implications:
{ First, our analysis revealed that when the interference power is time-varying due to channel fading, then for a given total average interference power, the performance of interfered wireless systems depends heavily on the structure of the interference rather than just the average interference power. 
In particular, characterizing the interference as constant noise leads to wrong performance assumptions of the system.}
Second, the mathematical treatment of wireless systems with interference channels that we propose here can serve as the basis for a system-level, cross-layer optimization.
Third, due to the network calculus approach used in this work, the obtained results can be easily extended to multi-hop settings, as we show in Section \ref{sec:SpecialCases}. 
{Finally,  the capacity expression for many other interesting channels, e.g., the secrecy channel, have similar structure to that of the interference channel. Hence, one can use our proposed approach and results 
to investigate the performance of such channels.}

The rest of the paper is structured as follows: In Section~\ref{sec:preliminaries}, we provide the required background on the stochastic network calculus.
The network calculus model of the interference channel is derived in Section~\ref{sec:derivations}.
Our numerical investigations are presented in Section~\ref{sec:evaluations}.
Finally, we provide brief conclusions in Section~\ref{sec:conclusions}.

%% file: Preliminaries.tex
\section{Stochastic Network Calculus for Wireless Channels}
\label{sec:preliminaries}
In this section,
we provide a brief description of the stochastic network calculus and its application to fading channels. We then adapt this network calculus to interference channels in the following section. The reader may refer to \cite{cruz, Book:chang, leboudec:networkcalculus, ciucu:networkservicecurvescaling2, fidler, snc_book, ciucu_schmitt, alzubaidy, fidler_tutorial} for more details on the network calculus and to \cite{jiang:servermodel, fidler_globecom, li:fading, mahmood:mimo, fidler:cde} for applications to wireless, fading, and Gilbert-Elliott channels.

\label{sec:snc_basics}

Stochastic network calculus considers queuing systems and networks of systems with stochastic arrival, departure, and service processes, where the bivariate functions $A(\tau,t)$, $D(\tau,t)$ and $S(\tau,t)$ for any $0 \le \tau \le t$ denote the \textit{cumulative} arrivals to the system, departures from the system, and service offered by the system, respectively, in the interval $[\tau,t)$.  We consider a discrete time model, where time-slots have a duration $T$ and $t \geq 0$ denotes the index of the respective time-slot.

A lossless system with service process $S(\tau,t)$
satisfies the input/output relationship
$D(0,t) \geq A \otimes S \left(0,t\right)$,
where $\otimes$ is the \minplus~convolution operator 
 given by $ x \otimes y \left(\tau,t\right) = \inf_{\tau \leq u \leq t} \left\{ x(\tau,u) + y (u,t) \right\}$.
A network service process of an $H$-hop path can be obtained using  the \minplus~convolution, i.e., $S_{\mathrm{net}} = S_1 \otimes  S_2 \otimes \dots \otimes S_H$.
In general, we are interested in probabilistic  bounds of the form
$\mathrm{Pr}\left[ W(t) > w^{\varepsilon} \right] \leq \varepsilon$, which is also known as the \textit{violation probability} for a target delay $w^{\varepsilon}$, under stable system operation.

Modeling wireless links in the context of network calculus however is  not  a trivial task. A particular difficulty arises when we seek to obtain a stochastic characterization of the cumulative service process of a wireless fading channel, as also witnessed in the context of the effective capacity of wireless systems \cite{wu}. A promising, recent approach for wireless networks has been proposed in \cite{alzubaidy} where the queuing behavior is analyzed directly in the ``domain'' of channel variations instead of the bit domain \cite{alzubaidy,AlZubaidyTON}. This can be interpreted as the \textit{SNR domain} (thinking of bits as ``SNR demands'' that reside in the system until these demands can be met by the channel).

To start with, the cumulative arrival, departure, and service processes in the bit domain, i.e., $A$, $D$, and $S$, are related to their SNR domain counterparts (represented in the following by calligraphic capital letters $\A$, $\D$, and $\S$) respectively, through the exponential function.
Thus, we have $\mathcal{A}(\tau,t) \triangleq e^{A(\tau,t)}$, $\mathcal{D}(\tau,t) \triangleq e^{D(\tau,t)}$, and $\mathcal{S}(\tau,t) \triangleq e^{S(\tau,t)}$.
Due to the exponential function, these cumulative functions become products of the increments in the bit domain. Assuming Shannon capacity
\begin{equation}
c_{t} = \log g\left(\gamma_t\right) = N \log_2\left( 1+ \gamma_t \right),
\label{eq:Shannon}
\end{equation}
where $c_t$ is the random service offered by the system in time-slot $t$, $N$ is the number of transmitted symbols per time-slot, and $\gamma_t$ is the instantaneous SNR, we obtain the cumulative service process in the SNR domain as
\begin{equation}
\label{eq:service_process_snr}
  \mathcal{S}(\tau,t)  = \prod_{u=\tau}^{t-1}  e^{c_u} = \prod_{u=\tau}^{t-1} g\left(\gamma_u\right) = \prod_{u=\tau}^{t-1} \left(1 + \gamma_u\right)^{\mathcal{N}},
\end{equation}
where $\mathcal{N} = N / \log 2$. To simplify notation, we consider the case $\mathcal{N} =1$ in the following. Performance bounds for the general case can be obtained by appropriately scaling the obtained results. Furthermore, in case of first-come first-served order, the delay at time $t$ is obtained
as follows
\begin{equation}
\label{eq:delay_snr}
  W(t)=\mathcal{W}(t)=\inf \{u \geq 0 : \A(0,t) / \D(0,t+u) \leq 1 \}.
\end{equation}
A bound $\varepsilon$ for the delay violation probability $\mathrm{Pr}\left[W(t)>w^\varepsilon\right]$ can be derived based on a transform of the cumulative arrival and service processes in the SNR domain using the moment bound.
In~\cite{alzubaidy}, it was shown that such a violation probability bound for a given $w^{\varepsilon}$ can be obtained as
\begin{equation}
\label{eq:delay_bound_snr}
  \varepsilon = \inf\limits_{s>0}\left\{\Kd(s, t+w^\varepsilon,t)\right\}.
\end{equation}
\noindent We refer to the function $\Kd\left(s,\tau,t\right)$ as the \textit{kernel} defined as
 \begin{equation}
\label{eq:function_M_Hussein}
  \Kd(s,\tau,t) = \sum_{u=0}^{\mathrm{min}(\tau,t)} \mathcal{M}_{\mathcal{A}}(1+s,u,t) \mathcal{M}_{\mathcal{S}}(1-s,u,\tau),
\end{equation}
where the function $\mathcal{M}_{\mathcal{X}}\left(s\right)$ is the Mellin transform \cite{Book:mellin} of a random process, defined as
\begin{equation}
\mathcal{M}_{\mathcal{X}}\left(s,\tau,t\right) = \mathcal{M}_{\mathcal{X}\left(\tau,t\right)} \left(s\right) = \mathrm{E}\left[\mathcal{X}^{s-1}\left(\tau,t\right)\right],
\label{eq:Mellin_Definition}
\end{equation}
for any $s \in \mathbb{C}$, where we restrict our derivations in this work to real valued $s \in \mathbb{R}$. We note that by definition of $\mathcal{X}(\tau,t) = e^{X(\tau,t)}$, the Mellin transform $\mathcal{M}_{\mathcal{X}}\left(s,\tau,t\right) = \mathrm{E}\left[e^{(s-1)X(\tau,t)}\right]$ after substitution of parameter $s = \theta+1$ implies also a solution for the moment-generating function (MGF), that is the basis of the effective capacity model \cite{wu} and of an MGF network calculus \cite{fidler}.

In the following, we will assume $\mathcal{A}\left(\tau,t\right)$ and $\mathcal{S}\left(\tau,t\right)$ to have stationary and independent increments.
We denote them by $\alpha$ for the arrivals (in SNR domain) and $g\left(\gamma\right)$ for the service.
Hence, the Mellin transforms become independent of the time instance, which we account for by denoting $\mathcal{M}_{\mathcal{X}}\left(s,\tau,t\right) = \mathcal{M}_{\mathcal{X}}\left(s, t - \tau\right)$.
In addition, as we only consider stable queuing systems in steady-state, the kernel becomes independent of the time instance $t$ and we denote $\Kd\left(s,t+w,t \right) \overset{t \to \infty}{=} \Kd\left(s,-w\right)$.

The strength of the Mellin-transform-based approach becomes apparent when considering block-fading channels.
The Mellin transform for the cumulative service process in SNR domain is given by
\begin{equation}
\label{eq:mellin_transform_service_basic}
 \mathcal{M}_{\mathcal{S}}\left(s,\tau,t\right)=\prod_{u=\tau}^{t-1} \mathcal{M}_{g(\gamma)}\left(s\right)=\mathcal{M}_{g(\gamma)}^{t-\tau}\left(s\right)= \mathcal{M}_{\S}\left(s,t-\tau\right)\, , \nonumber
\end{equation}
where $\mathcal{M}_{g(\gamma)}\left(s\right)$ is the Mellin transform of the stationary and independent service increment $g\left(\gamma\right)$ in the SNR domain.
The function $g\left(\cdot\right)$ represents here the modification of the SNR due to the Shannon formula Eq.~\eqref{eq:Shannon}. However, it can also model more complex system characteristics, most importantly scheduling effects.

Assuming the cumulative arrival process in SNR domain to have stationary and independent increments and denoting the corresponding Mellin transform by $\M_{\A}\left(s,t - \tau\right) =   \prod_{i=\tau}^{t-1}\M_{\alpha}(s)$, the steady-state kernel for a fading wireless channel is given by \cite{alzubaidy}
\begin{equation}
  \Kd\left(s,-w\right) = \frac{\M_{g\left(\gamma\right)}^w\left(1 - s\right)}{1 - \M_{\alpha}\left(1 + s \right) \M_{g\left(\gamma\right)}\left(1 - s \right)}
  \label{eq:delay_kernel}
\end{equation}
for any $s > 0$, under the stability condition
\begin{equation}
\label{eq:stability_cond}
  \M_{\alpha}\left(1 + s \right) \M_{g\left(\gamma\right)}\left(1 - s \right) < 1.
\end{equation}

Assuming Rayleigh fading, i.e., an exponentially distributed SNR with average $\gamma_{\emptyset}$ at the receiver, the Mellin transform results into \cite{alzubaidy}
\begin{equation}
\label{eq:service_process_mellin}
  \mathcal{M}_{g(\gamma)}\left(s\right)=e^{\frac{1}{\gamma_{\emptyset}}} {\gamma_{\emptyset}}^{s-1} \Gamma\left(s,{\gamma_{\emptyset}^{-1}}\right).
\end{equation}
\noindent where $\Gamma(x,y)=\int_y^{\infty} t^{x-1}e^{-t}\mathrm{d}t$ is the incomplete Gamma function.
Then the steady-state kernel for a Rayleigh-fading wireless channel turns out to be
\begin{multline}
  \Kd\left(s,-w\right) =
  \frac{\left(e^{\nicefrac{1}{\gamma_{\emptyset}}} {\gamma_{\emptyset}}^{-s } \Gamma\big(1-s,\frac{1}{\gamma_{\emptyset}}\big)\right)^{w}}{1- \M_{\alpha}\left(1 + s\right)  e^{\nicefrac{1}{\gamma_{\emptyset}}}  {\gamma_{\emptyset}}^{-s } \Gamma\big(1- s ,\frac{1}{\gamma_{\emptyset}}\big)}
  \label{eq:delay_kernel_rayleigh}
\end{multline}
for any $s > 0$ and under the stability condition in Eq.~(\ref{eq:stability_cond}). By substitution of the kernel Eq.~\eqref{eq:delay_kernel_rayleigh} into Eq.~\eqref{eq:delay_bound_snr}, a bound of the delay violation probability $\varepsilon$ for a given $w^{\varepsilon}$ can be obtained.

%% file: Derivations.tex
\section{Performance of Interference Channels}
\label{sec:derivations}

We first introduce a characterization of interference channels assuming independent block-fading processes for a transmitter/receiver pair and an arbitrary number of interferers. We then use this channel model to compute probabilistic performance bounds for interference channels in terms of their fading parameters.

\subsection{Block-Fading Interference Channel}
\label{sec:system_model}
Consider a wireless communication scenario with one transmitter/receiver pair that is subject to interfering signals from a set $\mathbf{I}$ of interferers.
Index $i \leq |\mathbf{I}|$ denotes the link between interferer $i$ and the receiver, while $i = 0$ denotes the link between the transmitter and the receiver (i.e., the signal of interest).
Denote by $P_{i}$ the transmit power per link, i.e., $P_{0}$ denotes the transmit power of the signal of interest, $P_{1}$ denotes the transmit power of interferer $1$, and so on.

The received power varies from time slot to time slot due to randomly varying channel gains of all links.
Denote the random channel gain of link $i$ during slot $t$ by $h_{i,t}$.
We focus in the following on random variations due to independent Rayleigh-distributed block-fading processes for all links $i$.
{ 
Hence, the received signal strength of link $i$ is given by $P_i |h_{i,t}|^2$ and is exponentially distributed with mean $p_{i} = P_{i} \cdot \mathrm{E}\left[|h_{i}|^2\right]$.
The fading is assumed to stay constant during one slot but varies independently from slot to slot.
Furthermore, the fading between different links is assumed to be statistically independent.
Based on the above definitions, the instantaneous signal-to-interference-plus-noise ratio (SINR) at the receiver during slot $t$ is a random variable and given as
\begin{equation}
\gamma_t = \frac{P_{0} |h_{0,t}|^2}{ \sum_{i} P_{i} |h_{i,t}|^2 +  \sigma^2} \; \mathrm{,}
\label{eqn:SINR}
\end{equation}     }
where $\sigma^2$ denotes the power of the additive white Gaussian noise (AWGN) process at the receiver.
Depending on the SINR, which is assumed to be known at the transmitter, the amount of information that can be conveyed changes in each time slot. 
In this work, we consider that for an SINR $\gamma_t$ the transmitter is able to transmit $c_t$ bits correctly to the receiver during slot $t$, where $c_t$ is defined by Eq. \eqref{eq:Shannon}.

\subsection{Derivation of $\M_{g(\gamma)}(s)$}
Based on the system model in Section~\ref{sec:system_model}, we proceed to present the main contributions of the paper.
Initially, we concentrate on deriving the Mellin transform of the service process for the interference channel. Then, we use the result to compute the kernel in Eq.~(\ref{eq:delay_kernel}).
Recall that we assume, apart from the signal of interest, $|\mathbf{I}|$ interferers to be present. The resulting SINR $\gamma_t$ is given by the ratio of exponentially distributed random variables in Eq.~\eqref{eqn:SINR}. Considering stationary and independent $\gamma_t$ for all $t$, we omit the index $t$. For $\gamma$ we have the distribution function \cite{Kandukuri_02}
\begin{equation}
\mathrm{Pr} \left[ \gamma \leq x \right] = F_{\gamma}\left(x\right) =  1 - e^{\frac{-x}{\gamma_{\emptyset}}}  \prod_{ \forall i \in \mathbf{I}} \frac{a_i}{a_i + x},
\label{eq:sir_dist}
\end{equation}
where $a_i = \frac{p_0}{p_i}$ denotes the ratio of the average received power from the signal of interest and interferer $i$.
Furthermore, $\gamma_{\emptyset} = \frac{p_0}{\sigma^2}$ is the noise-limited average SNR of the signal of interest.
Based on partial fractions decomposition \cite{Book:Polyanin}, it was shown \cite{parruca_2015} that the CDF in Eq.~\eqref{eq:sir_dist} can be reorganized as
\begin{equation}
F_{\gamma}\left(x\right) = 1 - e^{\frac{-x}{\gamma_{\emptyset}}} \sum_{\forall i \in \mathbf{I}} \frac{u_i}{a_i + x},
\label{eq:SINR_dist}
\end{equation}
where $u_i = \prod_{\forall s \in \mathbf{I}} a_s \left(\prod_{\forall t \in \mathbf{I} \setminus \{i\}} \left(a_t - a_i\right) \right) ^{-1}$ if there are multiple interferers $|\mathbf{I}| \geq 2$, and $u_1 = a_1$ if there is only one interferer $|\mathbf{I}| = 1$.
Note that the above representation only holds for interferers with different interference strengths, that is $a_i \neq a_j$ for all $i \neq j$.

First, we determine the Mellin transform for $g(\gamma)$, where $\gamma$ is the instantaneous SINR distributed  according to Eq.~(\ref{eq:SINR_dist}). The Mellin transform is given by
\begin{equation}
\mathcal{M}_{g(\gamma)} \left(s\right) = \mathrm{E}\left[ g(\gamma)^{s-1}\right] = \int_{0}^{\infty} (1+x)^{s-1} \, \mathrm{d}F_{\gamma}\left(x\right)
\label{eq:Mellin_Transform_SINR}
\end{equation}
for  $s < 1$.
To solve the integration above, we need the following lemma.

\begin{lemma}\label{lem:integral}
\begin{align*}
&  \int\limits_{0}^{\infty} \frac{(1+x)^{s-2}}{a_i + x} e^{\frac{-x}{\gamma_{\emptyset}}} \mathrm{d}x = e^{\frac{1}{\gamma_{\emptyset}}}\Big ( I^\infty_{1}(s) + I_{\delta}(s) + I^\infty_{2}(s)  \Big) \; \mathrm{,}
\end{align*}
for any small $\delta > 0$, where
\begin{align}
 I^k_{1}(s)\! = \!\! \sum\limits_{n=0}^{k} & \frac{(-1)^n \gamma_\emptyset^{s+n-1}}{(a_i-1)^{n+1}}  \Bigg[  \Gamma\left(s+n-1,\frac{1}{\gamma_\emptyset}\right)\notag \\
& - \Gamma\left(\!s\!+\!n\!-\!1,\frac{a_i\!-\!1 \!-\!\delta)}{\gamma_\emptyset}\!\right) \Bigg ]
 \; \mathrm{,}
\label{eq:difficult_guy1}
\end{align}
for $a_i > 2 + \delta$ and $I^k_{1}(s) = 0$, otherwise;
\begin{align}
& I_{\delta}(s) =   \int\limits_{\max(1,a_i-1-\delta)}^{\max(1,a_i-1+\delta)} \frac{z^{s-2}}{z + a_i-1} e^{\frac{-z}{\gamma_{\emptyset}}} \mathrm{d}z
\label{eq:difficult_guy2}
\end{align}
and
\begin{align}
I^k_{2}(s) &=\!
 \sum\limits_{n=0}^{k} \left( 1\!-\!a_i\right)^n  \gamma_{\emptyset}^{s-n-2} \Gamma\left(\! s\!-\!n\!-\!2, \frac{\max(1,a_i\!-\!1\!+\!\delta)}{\gamma_{\emptyset}}\! \right)
\label{eq:difficult_guy3}
\end{align}
for $k\ge 0$.
\end{lemma}

\begin{IEEEproof}
To solve the integral in Lemma \ref{lem:integral}, we start by performing a change of variable and letting $z=x+1$. Then,
\begin{align}
&  \int\limits_{0}^{\infty} \frac{(1+x)^{s-2}}{x+ a_i } e^{\frac{-x}{\gamma_{\emptyset}}} \mathrm{d}x = e^{\frac{1}{\gamma_{\emptyset}}} \int\limits_{1 }^{\infty} \! \frac{z^{s-2}}{z + a_i-1} e^{\frac{-z}{\gamma_{\emptyset}}} \mathrm{d}z.
\label{eq:difficult_guy4}
\end{align}
Note that since $a_i >0$ by definition and $z \in [1,\infty)$, we have $z+a_i-1 > 0$.

For the integral in the right hand side of Eq. (\ref{eq:difficult_guy4}), we use the following series representation
\begin{equation} \label{eq:series_expansion}
\frac{1}{b+1} = \sum_{n=0}^\infty (-1)^n b^n, \quad \text{for } \vert b \vert <1.
\end{equation}
To ensure that $\vert b \vert <1$, we partition the integral in Eq.~\eqref{eq:difficult_guy4} into three parts given as
\begin{equation}\label{eq:Int_Partition}
	\int\limits_{1}^{\infty}  f(z) \mathrm{d}z   =  \!\!\!\!\!\! \int\limits_{1}^{\max(1,a_i\!-\!1\!-\!\delta)} \!\!\!\!\!\!\!\!\! f(z) \mathrm{d}z + \!\!\!\!\!\! \int\limits_{\max(1,a_i\!-\!1\!-\!\delta)}^{\max(1,a_i\!-\!1\!+\!\delta)} \!\!\!\!\!\!\!\!\! f(z) \mathrm{d}z + \!\!\!\!\!\!  \int\limits_{\max(1,a_i\!-\!1\!+\!\delta)}^{\infty} \!\!\!\!\!\!\!\!\! f(z) \mathrm{d}z .
\end{equation}
In the expression above, the second term, which is Eq.~\eqref{eq:difficult_guy2}, diminishes when $\delta \rightarrow 0$. Depending on the range of values of $a_i$, we identify three cases:
(i) $a_i \in (0,2-\delta]$, then the first term and the second term will evaluate to zero and the integral in the third term will start at one; (ii) $a_i \in (2-\delta,2+\delta]$, then  the first term will again evaluate to zero, the integral in the second term will start at one, while the integral in the third term will start at $a_i-1+\delta$; and the last case (iii) $a_i \in (2+\delta,\infty)$, then the first term will have a value greater than zero, the integral in the second term will start at $z = a_i-1-\delta$ and the integral in the third term will start at $z = a_i-1+\delta$.

We can now apply the series expansion to the first and third term of Eq.~(\ref{eq:Int_Partition}) above. Starting with the integral in the first term, after multiplication with $a_i-1$ and considering the non-trivial case $a_i > 2+\delta$, we have
\begin{align}\label{eq:First_term}
& \int\limits_{1}^{a_i-1-\delta} \frac{z^{s-2}}{\frac{z}{a_i-1}+1}e^{-\frac{z}{\gamma_\emptyset}} \mathrm{d}z = \!\!\!
 \int\limits_{1}^{a_i-1-\delta} \!\!\sum\limits_{n=0}^{\infty} \left( \frac{-z}{a_i\!-\!1} \right)^n \!z^{s-2} e^{-\frac{z}{\gamma_\emptyset}} \mathrm{d}z  \nonumber \\
&=  \sum\limits_{n=0}^{\infty}   \frac{(-1)^n \gamma_\emptyset^{s+n-1} }{(a_i-1)^{n}} \int\limits_{1/\gamma_\emptyset}^{(a_i-1-\delta) / \gamma_\emptyset}  y^{s+n-2} e^{-y} \mathrm{d}y \nonumber \\
&= (a_i -1)I_1^\infty(s),
\end{align}
where in the second line we used the change of variables $y={\frac{z}{\gamma_\emptyset}}$. The last equality follows from the definition of the incomplete Gamma function.

For the third term in Eq.~\eqref{eq:Int_Partition}, we compute
\begin{align}
& \int\limits_{\max(1,a_i-1+\delta)}^{\infty} \frac{z^{s-3}}{ 1+\frac{a_i-1}{z}}e^{-\frac{z}{\gamma_\emptyset}} \mathrm{d}z \nonumber \\
&= \int\limits_{\max(1,a_i-1+\delta)}^{\infty} \sum\limits_{n=0}^{\infty} \left( \frac{1-a_i}{z} \right)^n \!z^{s-3} e^{-\frac{z}{\gamma_\emptyset}} \mathrm{d}z \nonumber \\
&=  \sum\limits_{n=0}^{k} \left( 1-a_i\right)^n  \gamma_{\emptyset}^{s-n-2}  \int\limits_{\max(1,a_i-1+\delta)/ \gamma_{\emptyset}}^{\infty}  y^{s-n-3}  e^{-y} \mathrm{d}y  \nonumber \\
&= I_2^\infty(s).
\label{eq:Third_term}
\end{align}
Again we used the change of variables $y={\frac{z}{\gamma_\emptyset}}$ in the third line and the definition of the incomplete Gamma function in the last line. The lemma follows by substituting Eq.~\eqref{eq:First_term} and Eq.~\eqref{eq:Third_term} in Eq.~\eqref{eq:Int_Partition} and by considering the range of $a_i$.
\end{IEEEproof}

In Lemma \ref{lem:integral}, we decompose the integral into three terms in order to be able to handle the evaluation of the otherwise intractable integral. The second term ($I_{\delta}(s)$) diminishes when $\delta$ approaches $0$ and may be ignored for $\delta \ll 1$. The first term ($I_{1}(s)$) has a value only when $a_i >2$, i.e., when $p_0> 2 p_i$ and is zero otherwise. The third term contributes to the solution for the entire range of $a_i$ and represents the complete solution when $a_i < 2$, i.e., $p_0 < 2p_i$.

\begin{theorem} \label{thm:1}
For the interference channel, we have
\begin{multline*}
\mathcal{M}_{g(\gamma)} \left(s\right) = \\ 1+ \sum_{\forall i \in \mathbf{I}} u_i \left ( (s-1) e^{\frac{1}{\gamma_{\emptyset}}}\Big( I^\infty_{1}(s) \!+\!\ I_{\delta}(s) \!+\! I^\infty_{2}(s)  \Big) \right),
\end{multline*}
for any $s<1$.
\end{theorem}

\begin{IEEEproof}
The Mellin transform of $g(\gamma)$ is given by  Eq.~\eqref{eq:Mellin_Transform_SINR} as
 \begin{equation}
\mathcal{M}_{g(\gamma)} \left(s\right)= \int_{0}^{\infty} (1+x)^{s-1} \, \mathrm{d}F_{\gamma}\left(x\right).
\label{eq:Mellin_Transform_SINR_2}
\end{equation}
Using integration by parts, we obtain for $s < 1$ that
\begin{align*}
\mathcal{M}_{g(\gamma)}(s) = & (1+x)^{s-1}  F_{\gamma}\left(x\right) \Big |_{0}^{\infty} \notag \\
& - (s-1) \int_{0}^{\infty} (1+x)^{s-2}  F_{\gamma}\left(x\right) \mathrm{d}x \notag \\
= &   - (s-1) \int_{0}^{\infty} (1+x)^{s-2}   \mathrm{d}x  \notag \\
& +  (s-1) \sum_{\forall i \in \mathbf{I}} u_i  \int\limits_{0}^{\infty} \frac{(1+x)^{s-2}}{a_i + x} e^{\frac{-x}{\gamma_{\emptyset}}} \mathrm{d}x   \notag \\
= &  1  +  (s-1) \sum_{\forall i \in \mathbf{I}} u_i  \int\limits_{0}^{\infty} \frac{(1+x)^{s-2}}{a_i + x} e^{\frac{-x}{\gamma_{\emptyset}}} \mathrm{d}x,
\end{align*}
where in the second step we inserted Eq.~\eqref{eq:SINR_dist}.
\end{IEEEproof}

Lemma \ref{lem:integral} gives the exact solution of the individual terms of Theorem~\ref{thm:1} expressed by infinite sums of incomplete Gamma functions. For numerical evaluation this may pose a computational problem. The following corollary provides easily computable bounds through truncation of these sums.

\begin{corollary}\label{cor:1}
For any even $k > 0$, $a_i > 1$, and $s<1$,
\begin{align}
& \Psi^{k+1}_i(s) \le \int\limits_{0}^{\infty} \frac{(1+x)^{s-2}}{a_i + x} e^{\frac{-x}{\gamma_{\emptyset}}} \mathrm{d}x \le \Psi^{k}_i(s) ,
\label{eq:corollary1}
\end{align}
where
\begin{align*}
& \Psi^{k}_i(s) = \lim_{\delta \to 0} e^{\frac{1}{\gamma_{\emptyset}}}\Big ( I^k_{1}(s) + I_{\delta}(s) + I^k_{2}(s)  \Big).
\end{align*}
The approximation error is bounded by $|\Psi^{k+1} - \Psi^{k}|$.
\end{corollary}
\begin{IEEEproof}
The proof follows directly from the monotonicity of the series expansion in Eq.~\eqref{eq:series_expansion} that has a limit of zero and the Leibniz alternating series test. The approximation error of the partial sum for $n=0, \ldots, k$, where $k$ is an even integer, is upper bounded by the $(k+1)^\mathrm{th}$ element of the series.
\end{IEEEproof}

The corollary above provides a practical way to bound the integral in Lemma~\ref{lem:integral}. In general, we are interested in an upper bound on the Mellin transform of the service $\M_{g(\gamma)}(s)$ for $s<1$ which provides an upper bound on the delay violation probability. Thus, truncation of the series at an even $k$ always leads to valid delay bounds.
The truncation error can be made arbitrarily small by choosing larger $k$. For the case $a_i < 1$ the series expansion of Lemma~\ref{lem:integral} uses the geometric series, where a truncation provides an approximate solution for the Mellin transform of the service. Again due to the convergence of the series, the truncation error can be made arbitrarily small by the choice of $k$.

Network layer performance bounds, e.g., a probabilistic delay bound, for a network of nodes with interference channels can be readily obtained from the Mellin transform of the service process of the channel which is characterized by Theorem~\ref{thm:1} and results from the network calculus presented in Section~\ref{sec:snc_basics}. The numerical results are presented in Section~\ref{sec:evaluations}.

\subsection{Asymptotes, Special Cases, and Multi-Hop Networks}\label{sec:SpecialCases}
In this subsection, we consider special cases, such as the noise-limited and the interference-limited channels. Also, we provide a solution of multi-hop networks.
\paragraph{Noise-Limited Channel}
In this case $p_0 \gg p_i$ so that $a_i = p_0/p_i \to \infty$. It follows that Eq.~\eqref{eq:sir_dist} evaluates to
\begin{equation*}
F_{\gamma}\left(x\right) =  1 - e^{\frac{-x}{\gamma_{\emptyset}}},
\end{equation*}
which is the CDF of the Rayleigh fading channel.

Considering Theorem~\ref{thm:1} and the case of a single interferer with $u_i = a_i$ the Mellin transform evaluates to
\begin{equation*}
\mathcal{M}_{g(\gamma)} \left(s\right) =  \left (1 + (s-1) e^{\frac{1}{\gamma_{\emptyset}}} a_i I^\infty_{1}(s) \right),
\end{equation*}
where we used that Eq.~\eqref{eq:difficult_guy2} and Eq.~\eqref{eq:Third_term} tend to zero for $a_i \to \infty$. The term $a_i I^\infty_{1}(s)$, where $I^\infty_{1}(s)$ is given by  Eq.~\eqref{eq:difficult_guy1}, evaluates for $a_i \to \infty$ to
\begin{equation*}
a_i I^\infty_{1}(s) = \gamma_\emptyset^{s-1} \Gamma\left(s-1,\frac{1}{\gamma_\emptyset}\right),
\end{equation*}
where we used that only the summand at $n=0$ does not tend to zero. The Mellin transform follows as
\begin{equation}
\mathcal{M}_{g(\gamma)} \left(s\right) = 1 + (s-1) e^{\frac{1}{\gamma_{\emptyset}}}\gamma_\emptyset^{s-1} \Gamma\left(s-1,\frac{1}{\gamma_\emptyset}\right).
\label{eq:rayleighalternative}
\end{equation}
Using the recurrence relation $\Gamma(s,x) = (s-1) \Gamma(s-1,x) + x^{s-1} e^{-x}$ we find that Eq.~\eqref{eq:rayleighalternative} is equivalent to the Mellin transform of the Rayleigh fading channel that was previously found in \cite{alzubaidy}, i.e., Eq.~\eqref{eq:service_process_mellin} is recovered.

\paragraph{Large Number of Interferers}
For a large number of interferers $|\mathbf{I}|$ with independent fading, the central limit theorem predicts that the combined interference power at the receiver becomes a Gaussian random variable.
Consequently, when the total power of the interference is split between infinitely many interferers, the variance of that random variable goes to zero and the total interference power at the receiver will have constant power $p_{\mathbf{I}}$. This has the same effect as an additional noise term with constant power $p_{\mathbf{I}}$, and thus the channel will behave like a noise-limited channel whose average SNR is equal to the average SINR $\frac{p_0}{\sigma^2 + p_{\mathbf{I}}}$ of the interference channel.

\paragraph{Identical Average Received Power}
We consider the special case where the average received power of the signal of interest and the signal of interferer $i$ are identical, i.e., $p_0 = p_i$ so that parameter $a_i$ evaluates to $a_i = 1$. By insertion of $a_i=1$ into Lemma~\ref{lem:integral}, we obtain $I_1^{\infty}(s) = 0$. Letting $\delta \to 0$, it follows that $I_{\delta}(s) \to 0$ diminishes. Finally, the first factor of $I_2^{\infty}$ given by Eq.~\eqref{eq:difficult_guy3}, i.e., $(1-a_i)^n$, is defined to be one for $n=0$ and zero otherwise, so that Lemma~\ref{lem:integral} evaluates to the simpler form
\begin{equation*}
\int\limits_{0}^{\infty} \frac{(1+x)^{s-2}}{a_i + x} e^{\frac{-x}{\gamma_{\emptyset}}} \mathrm{d}x = e^{\frac{1}{\gamma_{\emptyset}}} \gamma_{\emptyset}^{s-2} \Gamma\left(s-2,\frac{1}{\gamma_{\emptyset}}\right).
\end{equation*}
For this special case, the above integral can also be solved directly without using the series expansion of Lemma~\ref{lem:integral}, resulting in the same solution.

Considering a single interferer, the Mellin transform from Theorem~\ref{thm:1} becomes
\begin{equation}
\mathcal{M}_{g(\gamma)} \left(s\right) = 1 + (s-1) e^{\frac{1}{\gamma_{\emptyset}}} \gamma_{\emptyset}^{s-2} \Gamma\left(s-2,\frac{1}{\gamma_{\emptyset}}\right),
\label{eq:singleequalinterferer}
\end{equation}
which closely resembles the form of the Rayleigh fading channel in Eq.~\eqref{eq:rayleighalternative}.
\begin{figure}
\centering
\includegraphics [width=3.5in]{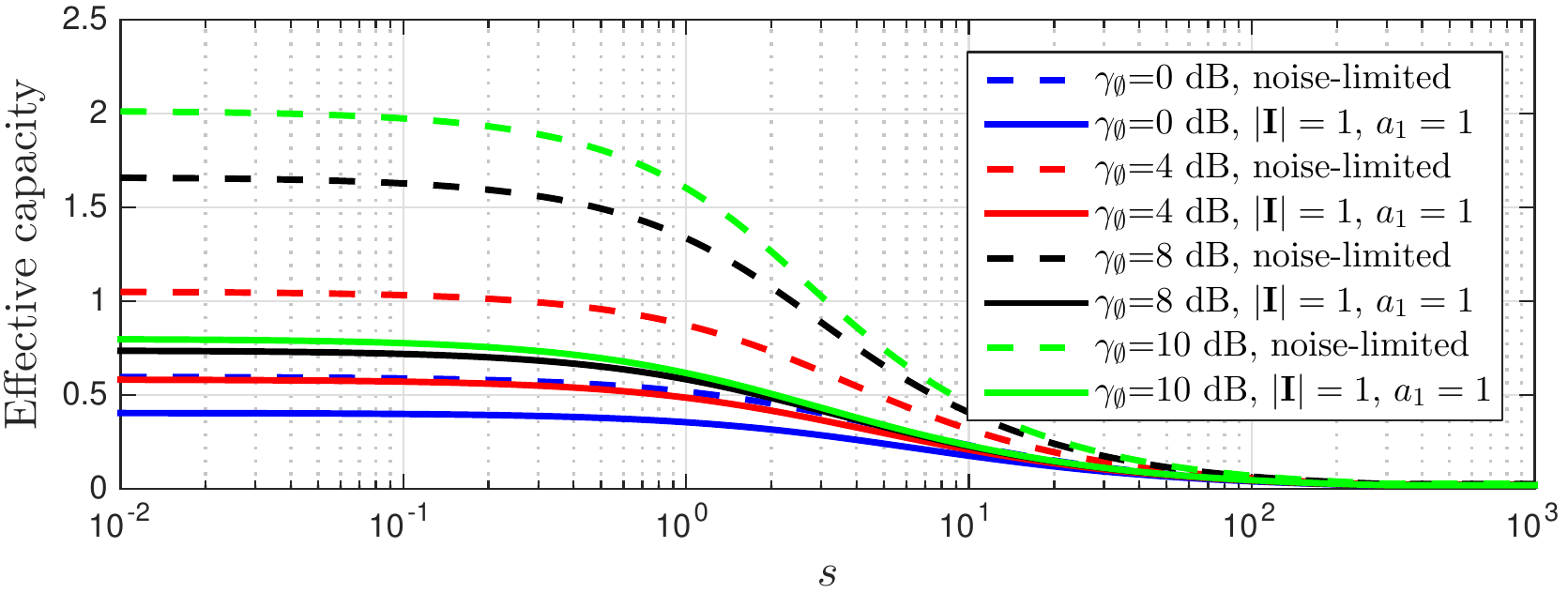}
\caption{Effective capacity (defined as $\frac{-1}{s} \log \M_{g(\gamma)} (1-s)$) as a function of $s$ for $\gamma_\emptyset \in\{0, 4, 8, 10\}$ dB, for an interference channel with a single interferer and identical average received power compared to noise-limited channel, assuming Rayleigh fading.}
\label{fig:1}
\vspace{3mm}
\end{figure}

Fig.~\ref{fig:1} depicts the effective capacity, computed as the normalized log Mellin transform of the service process defined as $\frac{-1}{s} \log \M_{g(\gamma)} (1-s)$, against parameter $s > 0$ for the special case (c) and compared to the noise-limited channel in case (a). 
As expected, when $s$ approaches zero, the effective capacity approaches the average channel capacity. As we increase $s$ (i.e., when a flow demands better QoS than mere average guarantees), the effective capacity diminishes and approaches the minimal capacity, that is zero. As expressed by Eqs.~\eqref{eq:rayleighalternative} and~\eqref{eq:singleequalinterferer}, the general shape of the effective capacity is the same for the case with and without interference. While the effective capacity of the noise-limited channel improves significantly as the channel SNR (i.e., $\gamma_\emptyset$) increases, the improvement is much smaller for the case with interference, and the effect of interference becomes more prominent with increasing SNR.

\paragraph{Interference-Limited Channel}
We characterize the interference-limited case by considering $\gamma_{\emptyset}\to \infty$, i.e., the noise power diminishes against the signal of interest as well as the interfering signals.
Consequently, the distribution in Eq.~\eqref{eq:SINR_dist} reduces to
\begin{equation}
F_{\gamma}\left(x\right) = 1 - \sum_{\forall i \in \mathbf{I}} \frac{u_i}{a_i + x}.
\label{eq:SINR_dist_int_limit}
\end{equation}
This leads to a structurally simpler solution. The exponential term in the integral of Lemma~\ref{lem:integral} disappears. The following lemma provides the solution for the new integral.
\begin{lemma}\label{lem:integral_int_case}
\begin{align*}
&  \int\limits_{0}^{\infty} \frac{(1+x)^{s-2}}{a_i + x} \mathrm{d}x =  \hat{I}^\infty_{1}(s) + \hat{I}_{\delta}(s) + \hat{I}^\infty_{2}(s),
\end{align*}
for any small $\delta > 0$, where
\begin{align}
\hat{I}^k_{1}(s)\! = \!\! \sum\limits_{n = 0}^{k} \frac{(-1)^n}{(a_i-1)^n} \left( \frac{\left(a_i - 1 - \delta\right)^{s - 1 +n} -1 }{s - 1 +n} \right),
\label{eq:difficult_guy1_int}
\end{align}
for $a_i > 2 + \delta$ and $\hat{I}^k_{1}(s) = 0$, otherwise;
\begin{align}
& \hat{I}_{\delta}(s) =   \int\limits_{\max(1,a_i-1-\delta)}^{\max(1,a_i-1+\delta)} \frac{z^{s-2}}{z + a_i-1} \mathrm{d}z
\label{eq:difficult_guy2_int}
\end{align}
and
\begin{align}
\hat{I}^k_{2}(s) &=\!
 \sum\limits_{n = 0}^{k}  \left(1 - a_i\right)^{n}  \frac{\left(a_i - 1 + \delta\right)^{s - 2 - n}}{s - 2 - n}
\label{eq:difficult_guy3_int}
\end{align}
for $k\ge 0$.
\end{lemma}
The proof follows very closely the one of Lemma~\ref{lem:integral} except that we consider Eq.~\eqref{eq:First_term} and Eq.~\eqref{eq:Third_term} with respect to the new distribution given in Eq.~\eqref{eq:SINR_dist_int_limit}.
We omit the detailed proof due to space constraints.
The corresponding Mellin transform can then be obtained by applying Theorem~\ref{thm:1} using  Lemma~\ref{lem:integral_int_case} instead of Lemma~\ref{lem:integral}.

\paragraph{Multi-Hop Interference Networks}
Stochastic network calculus allows the representation of a multi-hop network by a single network service process $\S_{\rm net}$, which is obtained by concatenating the service processes for all $H$ nodes along the traversed path, i.e.,
 \begin{equation}\label{eq:concatenationN}
\S_{\rm net}(\tau,t) =   \S_1  \conv \S_2  \conv \cdots \conv \S_H (\tau,t),
\end{equation}
where $\conv$ is the \mx~convolution operator defined in Section~\ref{sec:snc_basics}.

A bound on the convolution of two independent service processes $\S_1(\tau,t)$ and $\S_2(\tau,t)$ can be obtained using the Mellin transform, for any $s<1$, as
\begin{equation} \label{eq:convMT}
 \M_{ \S_1 \conv  \S_2} (s, \tau, t ) \le \sum_{u=\tau}^{ t}
\M_{ \S_1} (s, \tau,u ) \M_{\S_2} (s, u, t ).
\end{equation}
For a cascade of $H$ independent and identically distributed (i.i.d.) fading channels, we obtain for $s<1$ \cite{AlZubaidyTON},
\begin{align} \label{eq:MSnet}
\M_{\S_{\rm net}}(s,\tau,t) \le
\binom{H-1+t-\tau}{t-\tau} \Big(\M_{g(\gamma)}(s) \Big)^{t-\tau},
\end{align}
where $\M_{g(\gamma)}(s) $ is given by Theorem \ref{thm:1}.
When $a_i >1$ and to simplify the  computation of the desired performance bound, we use Corollary~\ref{cor:1} to bound
$\M_{g(\gamma)}(s) $, then Eq.~\eqref{eq:MSnet} becomes
\begin{align*}
\M_{\S_{\rm net}}(s,\tau,t)  \le &
\binom{H-1+t-\tau}{t-\tau} \\
& \cdot \left ( 1 + \sum_{\forall i \in \mathbf{I}} u_i \left ( (s-1)\Psi^{k}_i(s) \right ) \right )^{t-\tau},
\end{align*}
for any $s <1$. Substituting the above in Eq.~\eqref{eq:function_M_Hussein} gives the desired network performance bound.
For a cascade of channels that are independent but have different distributions, the joint Mellin transform can still be obtained from the individual ones, however the expressions are more complex \cite{Petreska_2015}.

%% file: Evaluations.tex
\section{Numerical Investigation}
\label{sec:evaluations}
In this section, we conduct numerical investigations of the interference channel performance based on the analysis presented in the previous section.
We first validate our analytical results using simulations.
Then we use the analytical model to address several important questions regarding the structure of the interference channel and its impact on the system performance.
{
In particular, we focus on the number of interferers as well as their absolute strengths, i.e., their average powers.
  We define the ratio of average received signal power to average  interference-plus-noise power at the receiver
  as
$\overline{\gamma} = \frac{p_0}{\sigma^2 + \sum_i p_i},$
which is equivalent to the average SINR of a system where the total interference power is  considered as an additive contribution to the noise.
}

In all of the following investigations, we choose the arrival model to be a constant rate process with rate $\rho$ measured in bits per time slot.
The Mellin transform of the arrival process is then given by $\M_{\A}\left(s,t - \tau\right) = e^{\rho(s-1)(t-\tau)}$. Using this Mellin transform and that of the service process derived in Section~\ref{sec:derivations} for the interference network, we obtain the kernel in Eq.~\eqref{eq:delay_kernel} and consequently the bound on the delay and violation probability based on Eq.~\eqref{eq:delay_bound_snr}.

\begin{figure}[t]
	\centering
	\subfloat[][]{
		\includegraphics[width=0.98\columnwidth]{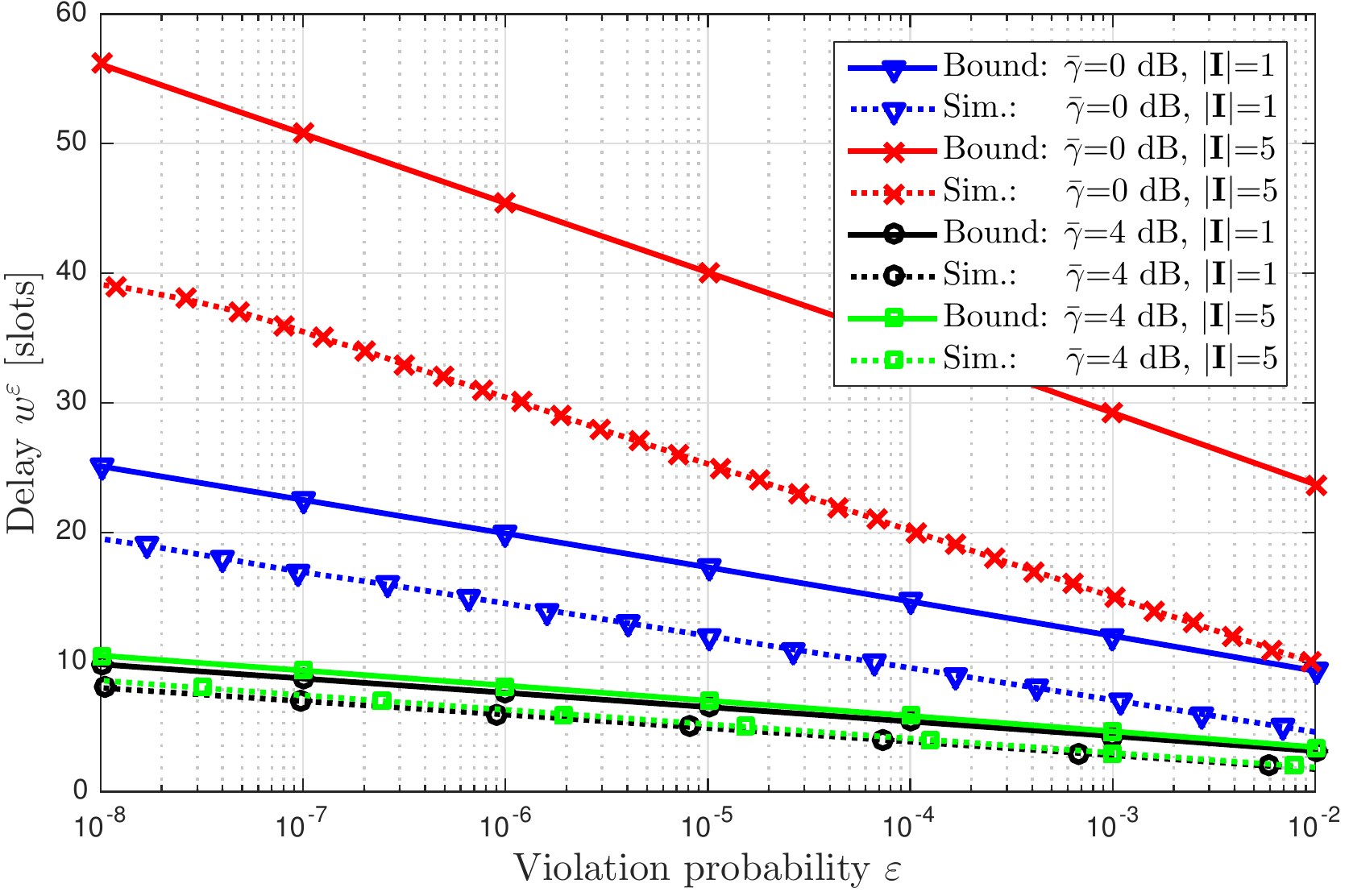}
		\label{fig:valid_delay_eps}
	}
	\hfil
	\subfloat[][]{
		\includegraphics[width=0.98\columnwidth]{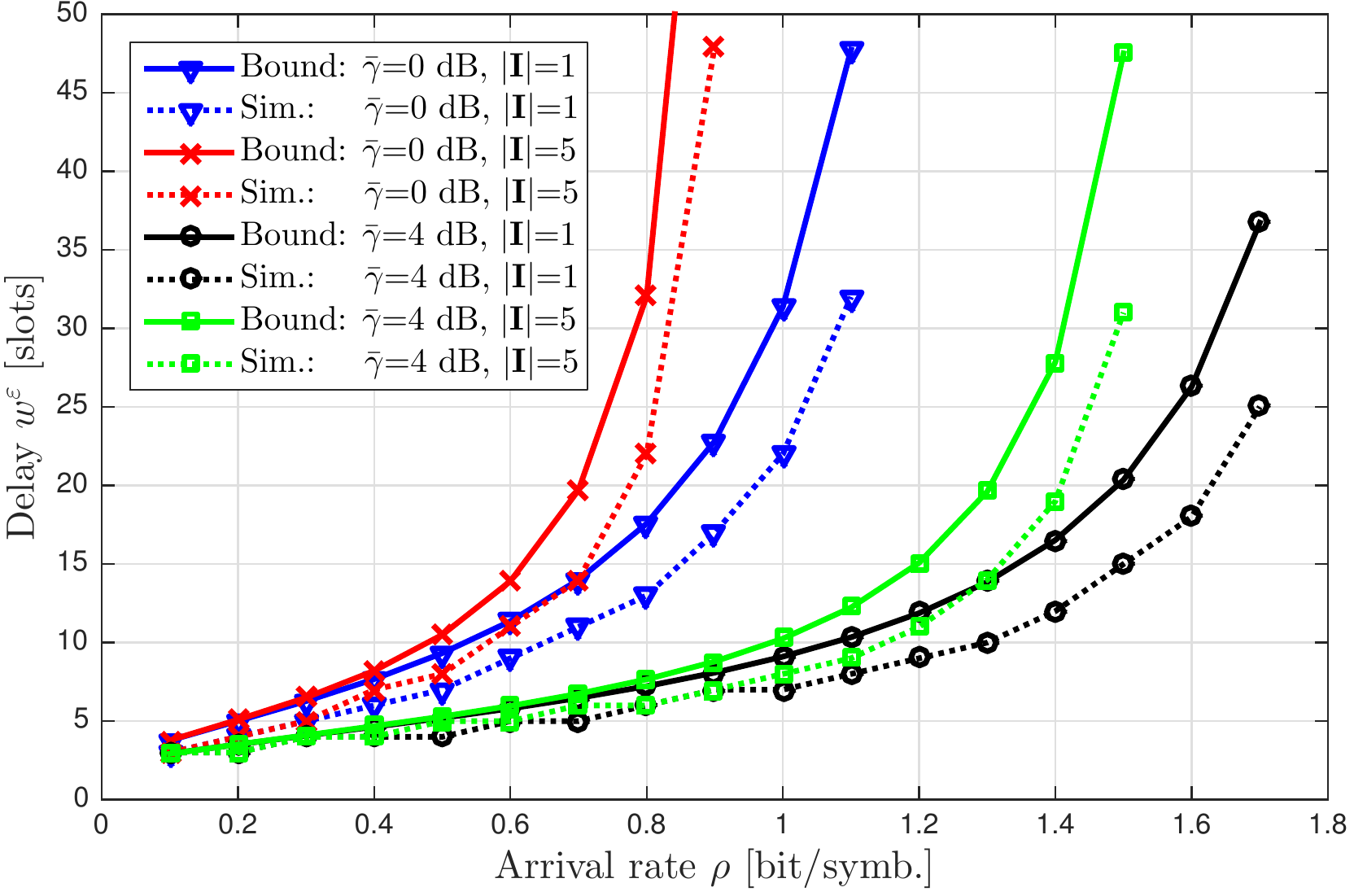}
		\label{fig:valid_delay_rate}
	}
	\caption{Validation of the analysis and delay performance with simulations for the parameter $\bar{\gamma}\in\{0,4\}$ and number of interferers $|\mathbf{I}|\in\{1,5\}$ with fixed average SNR $\gamma_\emptyset=15$ dB: $\mathrm{(a)}$ Delay bound ($w^\varepsilon$) in slots versus violation probability ($\varepsilon$) with fixed arrival rate $\rho=0.85$ bit/symbol. $\mathrm{(b)}$ Delay bound ($w^\varepsilon$) in slots versus arrival rate ($\rho$) in bit/symbol with fixed $\varepsilon=10^{-6}$.}
	\label{fig:validation}
	\vspace{4mm}
\end{figure}

\subsection{Validation of the Analytical Bounds}
{

We validate our computed bounds using simulation. We simulate a queuing system with service process given by the channel capacity of the interference channel, an arrival process with constant rate $\rho$ and with FIFO service discipline.  In order to estimate the target delay violation probabilities in the order of $10^{-6}$, we run the simulation for $10^{10}$ slots. 
In Fig.~\ref{fig:valid_delay_eps} we show the delay bound ($w^\varepsilon$) measured in transmission slots versus the violation probability ($\mathrm{Pr}[W(t)>w^\varepsilon]\le \varepsilon$) and compare it to the simulated delay.
We study the delay performance for different combinations of $\overline{\gamma}$ and number of interferers ($|\mathbf{I}|$), while we keep the arrival rate and the average SNR $\gamma_\emptyset=15$ dB  constant for all curves. These combinations reflect a wide delay performance range.
As expected, we observe  that the analytical results (solid curves) indeed are  upper bounds for the performance of their corresponding simulated systems (dotted curves).
Furthermore, we observe that the bounds are tight enough for a reasonable estimation of the system performance. It also shows that bounds grow tighter as the system becomes less utilized.
More importantly, in all cases the slope (i.e., the exponential decay) of the analytical and simulated curves match, therefore the relative gap diminishes as the delay grows larger.

In Fig.~\ref{fig:valid_delay_rate}, we show the delay bound ($w^\varepsilon$) measured in transmission slots versus the arrival rate for a violation probability of $\varepsilon = 10^{-6}$ and using the same parameter combinations as in Fig.~\ref{fig:valid_delay_eps}.
Again, we observe that the analytical results provide a reasonable bound for the simulated system performance.  The figure also shows that the bound accurately predicts the system stability region. Therefore, we conclude that $w^{\eps}$ is a reasonable upper bound for the system's delay performance.
In the rest of this section, we focus only on the analytical delay bounds to study different trends of the system performance.
}

\subsection{Effect of the Number of Interferers}
\begin{figure}[t]
\centering
\includegraphics[width=1\columnwidth]{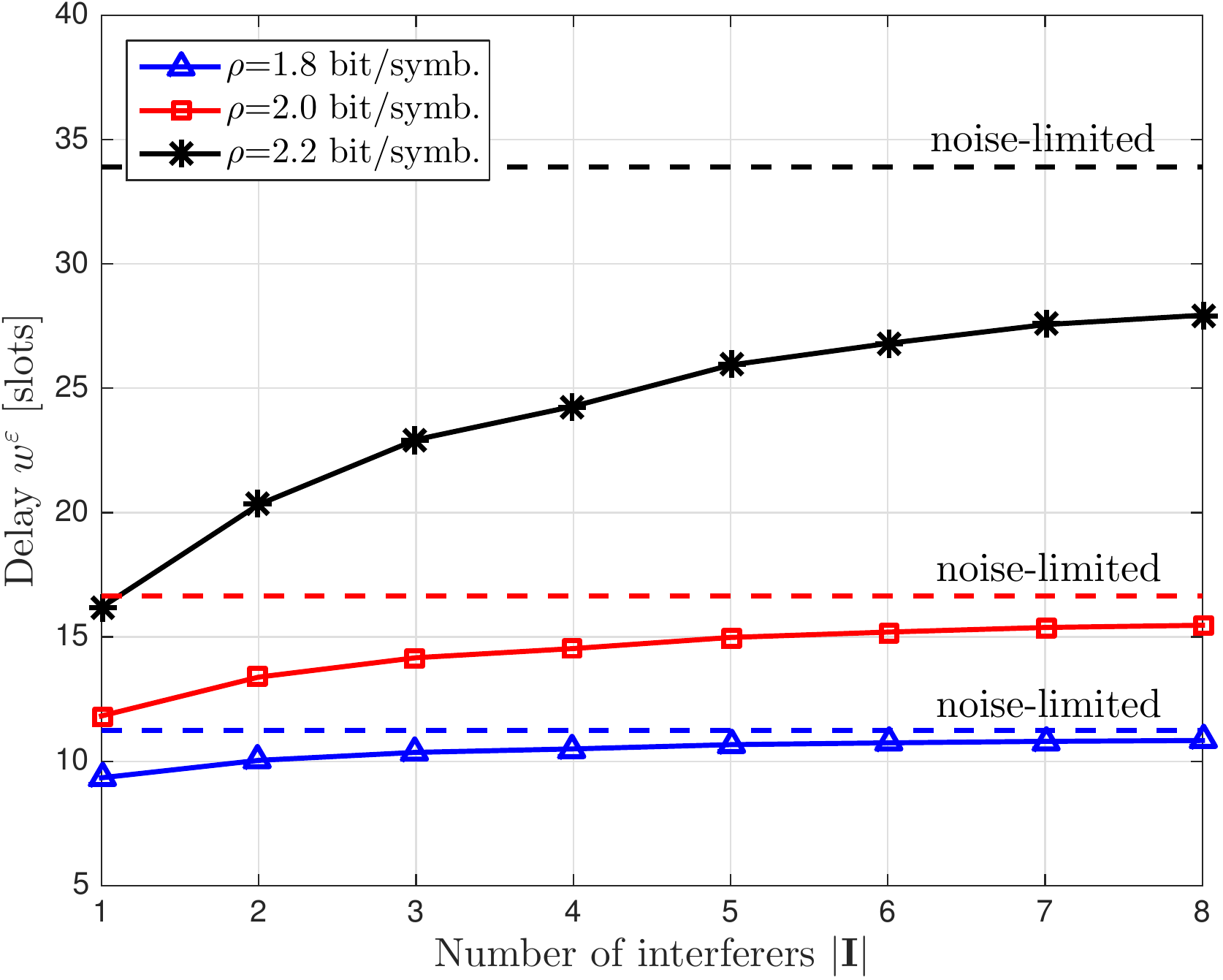}
\caption{Delay bound ($w^\varepsilon$) in slots versus the number of interferers ($|\mathbf{I}|$) for different arrival rates $\rho\in[1.8,2.0,2.2]$ with fixed $\bar{\gamma}=8$ dB, average SNR $\gamma_{\emptyset}=15$ dB, and $\varepsilon=10^{-6}$. The delay for the noise-limited case with average SNR $\gamma_\emptyset=8$ dB is also shown.}
\label{fig:delay_nI}
\vspace{3mm}
\end{figure}

{
Compared to state-of-the-art networking models that view interference as an additional constant contribution to the noise~\cite{Musavian_2010,Elalem_2013}, our explicit consideration of the individual random fading processes of the interferers enables us to address more fundamental questions. One interesting question is the following: Given an average total interference power, what is the  impact of the number of interferers on the system performance?

To answer the question above, we show in Fig.~\ref{fig:delay_nI} the delay as a function of the number of interferers ($|\mathbf{I}|$) in the network for different arrival rates. In this scenario, the average SNR is set to $\gamma_{\emptyset}=15$ dB and the delay violation probability to $\varepsilon=10^{-6}$.
We fix the parameter $\overline{\gamma}=8$ dB, i.e. the sum of the average interference powers stays constant. When there are multiple interferers, the total interference power is distributed among the interferers almost equally without violating the constraint $a_i\neq a_j$ in Eq.~\eqref{eq:SINR_dist}. However, when the number of interferers grows to infinity,  then the combined interference
can be modeled as additive noise with constant power as it was demonstrated in~\cite{Musavian_2010,Elalem_2013}.
This corresponds to the noise-limited case with average SNR $\gamma_\emptyset=8$ dB, which we also show for comparison.
}
The figure clearly reveals the importance of the structure of the interference that a certain transmitter/receiver pair is exposed to.
In general, if the sum of the average interference powers is kept constant, the more interferers are present, the worse the system performance gets. 
This happens because in the case of only few interferers, the variance of the SINR is higher, i.e., occasionally the interference is very small, leading to very high transmission rates. 
In contrast, the higher the number of interferers, the lower is the variance of the SINR, leading a decreasing performance.

\subsection{Effect of Main Signal Power}
{
Next, we study the effect of the SNR of the signal-of-interest on the system performance.
For a system that is operating at a fixed arrival rate and for a given number of interferers, we would like to study the effect of increasing the average signal strength and the interference strength simultaneously, such that the average signal to the average interference plus noise ($\overline{\gamma}$) remains constant.
}

\begin{figure}[t]
\centering
\subfloat[][]{
\includegraphics[width=1\columnwidth]{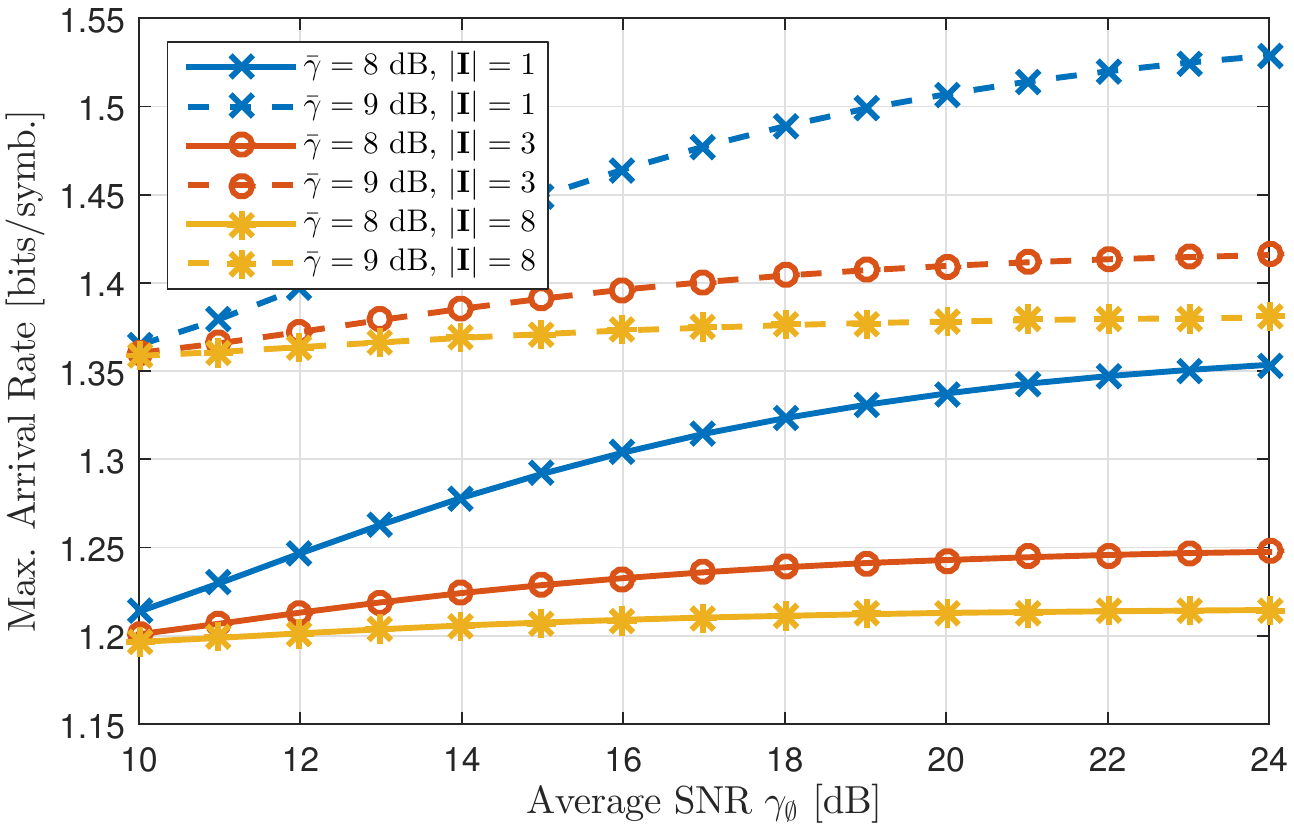}
\label{fig:maxrate_snr2}
}
\vspace{1mm}
\subfloat[][]{
\centering
\includegraphics[width=1\columnwidth]{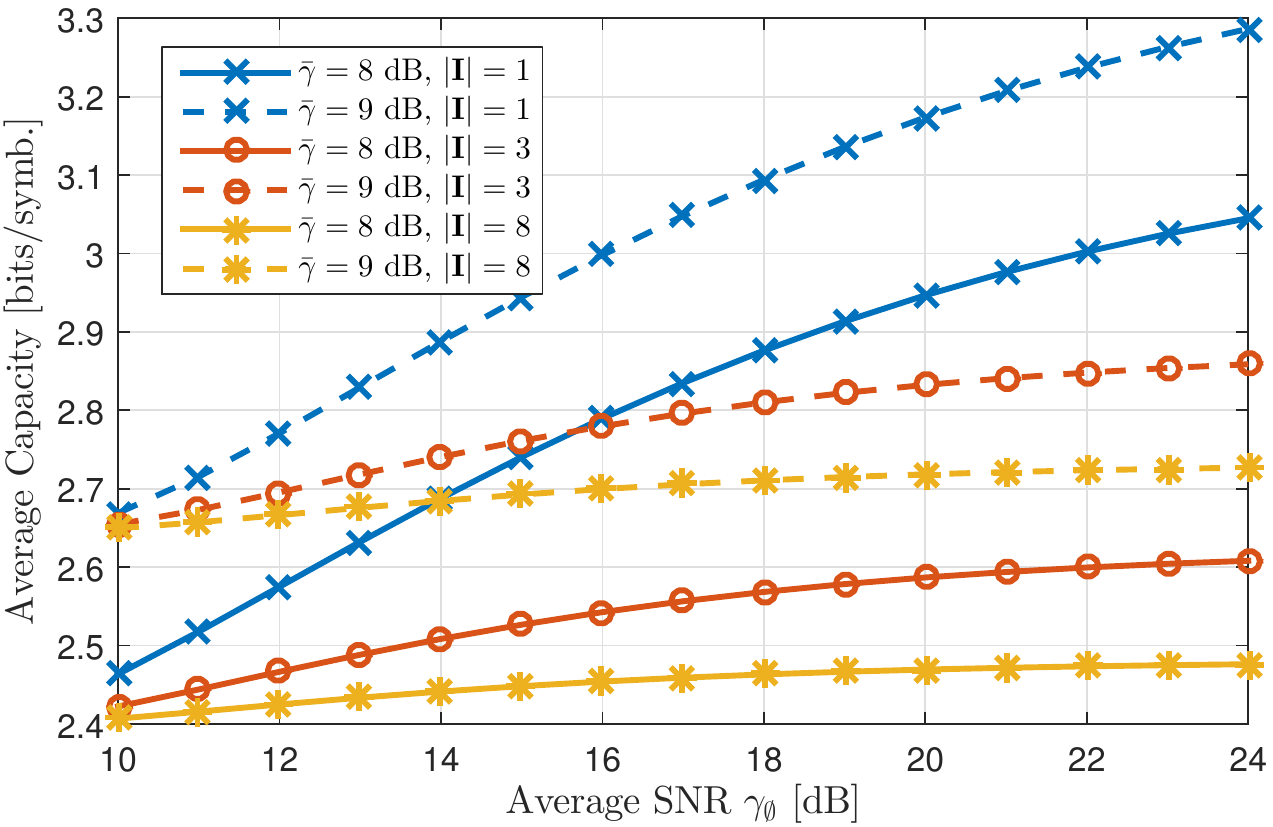}
\label{fig:avgcap_snr2}
}
\caption{Effect of average SNR $\gamma_\emptyset$ for different values of $\bar{\gamma} \in \{8,9\}$ and number of interferers $|\mathbf{I}|\in\{1,3,8\}$: (a) Maximum constant arrival rate versus average SNR ($\gamma_{\emptyset}$) so that delay bound still satisfies $w^\varepsilon=10$ and $\varepsilon=10^{-6}$. (b) Average capacity versus average SNR ($\gamma_{\emptyset}$) } 
\end{figure}

{

Fig.~\ref{fig:maxrate_snr2} confirms the observation that the performance decreases when there are more interferers, despite keeping the summed average interference power constant. Although this was already shown in Fig.~\ref{fig:delay_nI}, we now want to study how this depends on the average SNR ($\gamma_{\emptyset}$) of the basic signal-of-interest. When the average SNR is small, the disturbance comes mostly from the noise rather than from the interference, such that the number of interferers has little impact. At high SNR, the noise becomes relatively small, and the performance is limited by the interference. Fig.~\ref{fig:avgcap_snr2} indicates that the increase in performance for fewer interferers is most likely due to a large increase in the average capacity, which occurs because sometimes the interference-plus-noise can become close to zero, such that there is some probability that the SINR is extremely high. Under delay constraints, it seems that more interferers still decrease the system performance, as seen in Fig.~\ref{fig:maxrate_snr2}. However, when comparing both figures at high SNR $\gamma_{\emptyset}$ for different values of $\bar{\gamma}$, it can be seen that higher average capacity does not always mean better performance under delay constraints. Fewer interferers lead to a large increase in average capacity, but only to a small or moderate increase in performance under delay constraints.

}

%% file: Conclusions.tex
\section{Conclusions}
\label{sec:conclusions}
In this paper, we considered interference channels, where the signal of interest and the signals of an arbitrary number of interferers experience independent Rayleigh fading. 
We provided a fundamental stochastic characterization of the time-varying channel capacity by its Mellin transform. 
Using the transform domain and network calculus queuing relations, we contributed the first higher layer performance evaluation of such channels which enabled us to reveal key aspects of interference channels. 
We showed that even for a fixed summed interference power, the interference channel has relevant degrees of freedom that impact the delay performance significantly, namely strength and number of interfering transmitters.
While our evaluations have shown this result for scenarios where the average sum interference power has been kept constant, similar conclusions can be drawn if the average SINR of the scenario is kept constant. 
Even in this case, the structure of the interference has a significant impact on the performance of the system.
As future work, we consider in particular the coupling of wireless systems as key next step that result from the work presented in this paper.

%% file: Interference_Main.bbl
\begin{thebibliography}{10}
\providecommand{\url}[1]{#1}
\csname url@samestyle\endcsname
\providecommand{\newblock}{\relax}
\providecommand{\bibinfo}[2]{#2}
\providecommand{\BIBentrySTDinterwordspacing}{\spaceskip=0pt\relax}
\providecommand{\BIBentryALTinterwordstretchfactor}{4}
\providecommand{\BIBentryALTinterwordspacing}{\spaceskip=\fontdimen2\font plus
\BIBentryALTinterwordstretchfactor\fontdimen3\font minus
  \fontdimen4\font\relax}
\providecommand{\BIBforeignlanguage}[2]{{%
\expandafter\ifx\csname l@#1\endcsname\relax
\typeout{** WARNING: IEEEtran.bst: No hyphenation pattern has been}%
\typeout{** loaded for the language `#1'. Using the pattern for}%
\typeout{** the default language instead.}%
\else
\language=\csname l@#1\endcsname
\fi
#2}}
\providecommand{\BIBdecl}{\relax}
\BIBdecl

\bibitem{Mass_Consult_09}
\BIBentryALTinterwordspacing
{Mass Consultants Ltd.}, ``Estimating the utilisation of key license-exempt
  spectrum bands.''\hskip 1em plus 0.5em minus 0.4em\relax Ofcom, 2009.
  [Online]. Available:
  \url{http://stakeholders.ofcom.org.uk/binaries/research/technology-research/wfiutilisation.pdf}
\BIBentrySTDinterwordspacing

\bibitem{Cadambe_2008}
V.~Cadambe and S.~Jafar, ``Interference alignment and degrees of freedom of the
  k-user interference channel,'' \emph{IEEE Trans. Inf. Theory}, vol.~54,
  no.~8, pp. 3425--3441, Aug. 2008.

\bibitem{Gesbert_2010}
D.~Gesbert, S.~Hanly, H.~Huang, S.~Shamai, O.~Simeone, and Y.~Wei, ``Multi-cell
  {MIMO} cooperative networks: A new look at interference,'' \emph{IEEE J. Sel
  Areas on Commun.}, vol.~28, no.~9, pp. 1380--1408, Dec. 2010.

\bibitem{Zhang_2013}
B.~Zhang, W.~Cheng, L.~Sun, X.~Cheng, T.~Znati, M.~A. Al-Rodhaan, and
  A.~Al-Dhelaan, ``Queuing modeling for delay analysis in mission oriented
  sensor networks under the protocol interference model,'' in \emph{Proc. {ACM}
  {MiSeNet}}, 2013, pp. 11--20.

\bibitem{Bisnik2009}
N.~Bisnik and A.~A. Abouzeid, ``Queuing network models for delay analysis of
  multihop wireless {Ad~Hoc} networks,'' \emph{Ad Hoc Networks}, vol.~7, no.~1,
  pp. 79--97, 2009.

\bibitem{Musavian_2010}
L.~Musavian, S.~Aissa, and S.~Lambotharan, ``Effective capacity for
  interference and delay constrained cognitive radio relay channels,''
  \emph{IEEE Trans. on Wireless Commun.}, vol.~9, no.~5, pp. 1698--1707, May
  2010.

\bibitem{Elalem_2013}
M.~Elalem and L.~Zhao, ``Effective capacity and interference analysis in
  multiband dynamic spectrum sensing,'' \emph{Communications and Network},
  vol.~5, no.~2, pp. 111--118, May 2013.

\bibitem{Le_10}
L.~B. Le, E.~Modiano, C.~Joo, and N.~B. Shroff, ``Longest-queue-first
  scheduling under {SINR} interference model,'' in \emph{Proc. ACM MobiHoc},
  Sept. 2010, pp. 41--50.

\bibitem{parruca_2015}
D.~Parruca and J.~Gross, ``On the interference as noise approximation in
  ofdma/lte networks,'' in \emph{Proc. {IEEE} {ICC}}, Jun. 2014.

\bibitem{alzubaidy}
H.~Al-Zubaidy, J.~Liebeherr, and A.~Burchard, ``{A (min, $\times$) Network
  Calculus for Multi-Hop Fading Channels},'' in \emph{Proc. IEEE INFOCOM},
  2013, pp. 1833--1841.

\bibitem{cruz}
R.~L. Cruz, ``A calculus for network delay. {I}. network elements in
  isolation,'' \emph{IEEE Trans. Inf. Theory}, vol.~37, no.~1, pp. 114--131,
  Jan. 1991.

\bibitem{Book:chang}
C.-S. Chang, \emph{Performance Guarantees in Communication Networks}.\hskip 1em
  plus 0.5em minus 0.4em\relax Springer-Verlag, 2000.

\bibitem{leboudec:networkcalculus}
J.-Y. {Le Boudec} and P.~Thiran, \emph{Network Calculus. A Theory of
  Deterministic Queuing Systems for the {I}nternet}.\hskip 1em plus 0.5em minus
  0.4em\relax Springer-Verlag, 2001.

\bibitem{ciucu:networkservicecurvescaling2}
F.~Ciucu, A.~Burchard, and J.~Liebeherr, ``Scaling properties of statistical
  end-to-end bounds in the network calculus,'' \emph{IEEE Trans. Inf. Theory},
  vol.~52, no.~6, pp. 2300--2312, Jun. 2006.

\bibitem{fidler}
M.~Fidler, ``An end-to-end {p}robabilistic {n}etwork {c}alculus with {m}oment
  {g}enerating {f}unctions,'' in \emph{Proc. of IEEE IWQoS}, Jun. 2006, pp.
  261--270.

\bibitem{snc_book}
Y.~Jiang and Y.~Liu, \emph{Stochastic Network Calculus}.\hskip 1em plus 0.5em
  minus 0.4em\relax Springer, 2008.

\bibitem{ciucu_schmitt}
F.~Ciucu and J.~Schmitt, ``Perspectives on network calculus - no free lunch but
  still good value,'' in \emph{Proc. {ACM} {SIGCOMM}}, Aug. 2012, pp. 311--322.

\bibitem{fidler_tutorial}
M.~Fidler and A.~Rizk, ``A guide to the stochastic network calculus,''
  \emph{IEEE Commun. Surveys Tuts.}, vol.~17, no.~1, pp. 92 -- 105, Mar. 2015.

\bibitem{jiang:servermodel}
Y.~Jiang and P.~J. Emstad, ``Analysis of stochastic service guarantees in
  communication networks: A server model,'' in \emph{Proc. {IEEE} {IWQoS}},
  Jun. 2005, pp. 233--245.

\bibitem{fidler_globecom}
M.~Fidler, ``A network calculus approach to probabilistic quality of service
  analysis of fading channels,'' in \emph{Proc. IEEE GLOBECOM}, Nov. 2006, pp.
  1--6.

\bibitem{li:fading}
C.~Li, H.~Che, and S.~Li, ``A wireless channel capacity model for quality of
  service,'' \emph{{IEEE} Trans. Wireless Commun.}, vol.~6, no.~1, pp.
  356--366, Jan. 2007.

\bibitem{mahmood:mimo}
K.~Mahmood, A.~Rizk, and Y.~Jiang, ``On the flow-level delay of a spatial
  multiplexing {MIMO} wireless channel,'' in \emph{Proc. of IEEE ICC}, Jun.
  2011.

\bibitem{fidler:cde}
M.~Fidler, R.~L\"ubben, and N.~Becker, ``Capacity-delay-error-boundaries: A
  composable model of sources and systems,'' \emph{{IEEE} Trans. Wireless
  Commun.}, vol.~14, no.~3, pp. 1280--1294, Mar. 2015.

\bibitem{wu}
D.~Wu and R.~Negi, ``Effective {c}apacity: {a} {w}ireless {l}ink {m}odel for
  {s}upport of {q}uality of {s}ervice,'' \emph{IEEE Trans. Wireless Commun.},
  vol.~2, no.~4, pp. 630--643, Jul. 2003.

\bibitem{AlZubaidyTON}
H.~Al-Zubaidy, J.~Liebeherr, and A.~Burchard, ``Network-layer performance
  analysis of multihop fading channels,'' \emph{IEEE/ACM Trans. Netw.},
  vol.~24, no.~1, pp. 204--217, Feb. 2016.

\bibitem{Book:mellin}
B.~Davies, \emph{Integral Transforms and Their Applications}.\hskip 1em plus
  0.5em minus 0.4em\relax Springer-Verlag, 1978.

\bibitem{Kandukuri_02}
S.~Kandukuri and S.~Boyd, ``Optimal power control in interference-limited
  fading wireless channels with outage-probability specifications,'' \emph{IEEE
  Trans. Wireless Commun.}, vol.~1, no.~1, pp. 46--55, Jan. 2002.

\bibitem{Book:Polyanin}
A.~Polyanin and A.~Manzhirov, \emph{Handbook of Mathematics for Engineers and
  Scientists}.\hskip 1em plus 0.5em minus 0.4em\relax CRC Press, 2006.

\bibitem{Petreska_2015}
N.~Petreska, H.~Zubaidy, R.~Knorr, and J.~Gross, ``On the recursive nature of
  end-to-end delay bound for heterogeneous wireless networks,'' in \emph{Proc.
  {IEEE} {ICC}}, Jun. 2015.

\end{thebibliography}
